\documentclass[times,onecolumn]{aastex631}

\usepackage{natbib}
\usepackage{amsmath}
\usepackage{amssymb}
\usepackage{subfigure}
\usepackage{url}
\usepackage{CJK}
\usepackage{longtable}
\usepackage{graphicx}   
\usepackage{epsfig}
\usepackage{bm}
\usepackage{multirow}

\usepackage{color}  
\usepackage{times}

\usepackage{booktabs}  

\renewcommand{\vec}[1]{ {\mathbf #1} }

\newcommand{\RNum}[1]{\uppercase\expandafter{\romannumeral #1\relax}}

\graphicspath{{figures/}}

\shorttitle{Threshold of Magnetic Quantities at Eruption Onset}
\shortauthors{Bian et al.}
\graphicspath{{figures/}}

\begin{document}

\title{An investigation of magnetic energy and helicity thresholds at the onset of solar eruptions based on numerical simulations}


\correspondingauthor{Chaowei Jiang} \email{chaowei@hit.edu.cn}

\author[0000-0001-9189-1846]{Xinkai Bian}
\affiliation{State Key Laboratory of Solar Activity and Space Weather, School of Aerospace, Harbin Institute of Technology, Shenzhen 518055, China}
\affiliation{School of Science, Harbin Institute of Technology, Shenzhen 518055, China}

\author[0000-0002-7018-6862]{Chaowei Jiang}
\affiliation{State Key Laboratory of Solar Activity and Space Weather, School of Aerospace, Harbin Institute of Technology, Shenzhen 518055, China}

\author{Qingjun Liu}
\affiliation{School of Opto-electronic Engineering, Zaozhuang University, Zaozhuang 277160, China}

\author[0000-0002-4581-9242]{Yang Wang}
\affiliation{School of Science, Harbin Institute of Technology, Shenzhen 518055, China}

\author[0000-0002-8474-0553]{Peng Zou}
\affiliation{State Key Laboratory of Solar Activity and Space Weather, School of Aerospace, Harbin Institute of Technology, Shenzhen 518055, China}

\author[0000-0001-8605-2159]{Xueshang Feng}
\affiliation{State Key Laboratory of Solar Activity and Space Weather, School of Aerospace, Harbin Institute of Technology, Shenzhen 518055, China}

\author[0000-0003-4711-0306]{Pingbing Zuo}
\affiliation{State Key Laboratory of Solar Activity and Space Weather, School of Aerospace, Harbin Institute of Technology, Shenzhen 518055, China}

\author[0000-0002-7094-9794]{Yi Wang}
\affiliation{State Key Laboratory of Solar Activity and Space Weather, School of Aerospace, Harbin Institute of Technology, Shenzhen 518055, China}

\begin{abstract}

Identifying universal, topology-independent thresholds in the coronal magnetic fields at onset of solar eruptions is crucial for physics-based prediction of eruptions. To this end, we systematically analyze the evolution of magnetic energy and helicity in twelve high-fidelity 3D magnetohydrodynamic simulations where eruptions are triggered by magnetic reconnection. 
The simulations encompass a comprehensive parameter space, including bipolar and quadrupolar configurations, sheared arcades and pre-existing flux ropes, and various photospheric driving motions. We find that the ratio of current-carrying helicity to total relative helicity $(H_j/H_r)$ exhibits a remarkably consistent threshold of $0.38 \pm 0.04$ at eruption onset across all cases, with a coefficient of variation of only $\sim 10$\%. This threshold specifically characterizes the critical conditions at eruption onset and is largely independent of the subsequent temporal evolution, making it the most robust eruptivity indicator identified. In contrast, other normalized helicity and energy metrics show greater scatter. Crucially, we further find that $H_j/H_r$ does not necessarily achieve its peak at the eruption onset time and its post-eruption evolution diverges based on magnetic topology: it continues to increase in bipolar configurations due to tether-cutting reconnection, which transforms sheared arcade into the erupting current-carrying magnetic flux, but decreases in quadrupolar configurations as breakout reconnection peels off the erupting flux. These results highlight the helicity ratio as a promising and consistent eruptivity indicator and provide new insights into its dynamic evolution due to different reconnections.

\end{abstract}


\section{Introduction}\label{sec:intro}

Coronal mass ejections (CMEs) are spectacular solar eruptions, and if directed to the Earth, they can induce severe disturbances in the geospace environment which are known as space weather. The initiation of CMEs are fundamentally governed by the coronal magnetic field, which stores free magnetic energy prior to the eruption and release it during the event. A central question in recent research of solar eruptions as well as space weather is what physical properties of the coronal magnetic field determine its eruptivity, based on which whether there exists some relevant parameters that can be used for eruption prediction. For this purpose, it is attractive to find some global scalar parameters without the need of analyzing specific magnetic configurations or the detailed mechanisms of eruptions.

Two fundamental quantities are unquestionably related to the eruptivity of coronal magnetic fields, i.e., the non-potentiality, which is the degree of deviation of magnetic field from the lowest energy state of potential field (e.g., \citealp{Jing2010, Gupta2021, Li2021}), and the topological complexity, which can be measured by relative magnetic helicity (e.g., \citealp{Zuccarello2018, Thalmann2019a, Gupta2021, Liu2023a}). 
Indeed, numerous studies have shown that major eruptions are associated with sufficiently large free magnetic energy~\citep{Moore2012, Sun2015}, and that active regions with higher helicity tend to exhibit stronger eruptivity \citep{Nindos2004, LaBonte2007, Smyrli2010, Tziotziou2012}. However, when these quantities are expressed in their extensive form, their absolute values depend on the size of the active region or the total unsigned magnetic flux on the photosphere. As a result, their threshold values vary significantly from event to event, which limits their applicability in statistical analyses of solar eruptions. To overcome this limitation and enable meaningful comparisons among eruptions associated with different magnetic flux contents, it is preferable to use intensive forms such as normalized or dimensionless parameters. The frequently used intensive parameters include the normalized free magnetic energy (i.e., ratio of free energy to potential field energy) and normalized helicity (i.e., ratio of relative helicity to square of total unsigned magnetic flux).

Since the three-dimensional (3D) coronal magnetic field is difficult to measure, some studies attempt to derive the magnetic energy and helicity contents in active regions from time sequences of photospheric magnetograms by computing the accumulated magnetic energy and helicity injected through the photosphere \citep{Leka2007, Li2022, Liokati2023, Sun2024, Li2025d, Sun2025}.
However, the photospheric plasma flow field, which is essential in computing the energy and helicity fluxes, is indirectly derived with large uncertainties. 
Other studies employ nonlinear force-free field (NLFFF) extrapolation methods to reconstruct the 3D coronal magnetic field~\citep{Grad1958, Wiegelmann2004, Jiang2011, Liu2025}, based on which the energy and helicity can be obtained by volume integration~\citep{Aschwanden2014, Thalmann2019, Gupta2021, Duan2023a, Liu2023a, Wang2023c}. For instance, \citet{Gupta2021} and \citet{Duan2023a} demonstrated that normalized free magnetic energy and normalized current-carrying helicity (which is the helicity only related to the magnetic field induced by the coronal currents) can effectively distinguish flares associated with CMEs from confined flares. 
However, NLFFF extrapolation methods are limited to static snapshots of the magnetic field and cannot follow its continuous dynamic evolution, making it difficult to track the evolution of these parameters during the eruption onset. Moreover, the extrapolated results may depend on the adopted numerical method and parameters choices~\citep{Wiegelmann2008}, which can lead to differences in the derived magnetic configurations and associated quantities. 

Numerical magnetohydrodynamic (MHD) simulations that provide complete access to the evolving 3D magnetic fields can guide the search for possible parameters as indicators of eruption onset. By comparing numerical experiments that simulate eruption initiations in different magnetic environments or different photospheric motions, it is helpful to identify which of the proposed parameters is useful to predict an eruption.
For example, \citet{Pariat2017} investigated different parameters associated with magnetic energy and helicity in the simulations of magnetic flux emergence into different background fields performed by \citet{Leake2013} and \citet{Leake2014}. They found for the first time that, compared to other parameters, only the ratio of current-carrying helicity to relative helicity (hereafter referred to as the helicity ratio) serves as a reliable indicator of eruptivity, since when an eruption is triggered, this helicity ratio is significantly higher than the non-eruptive cases. 
Further, \citet{Pariat2023} confirmed this finding by comparing the helicity ratio in two similar simulations performed by \citet{Wyper2017} in which one produced a coronal jet and the other not. They showed that the helicity ratio serves as a discriminating metric, it increases to a peak value of about $0.8$ prior to jet onset and subsequently decreases in the jet-producing case, whereas it remains nearly constant in the non-eruptive case.
Similarly, \citet{Zuccarello2018} analyzed the flux rope formation and eruption simulations presented in \citet{Zuccarello2015}, in which a flux rope gradually formed through a range of different boundary driven motions and subsequently erupted via the torus instability~\citep{Kliem2006, Aulanier2010, Demoulin2010, Kliem2014}.
They found that in all the cases, the helicity ratio consistently converges to $\sim 0.3$ before eruption, suggesting that the helicity ratio can serve as a proxy of the onset criterion of torus instability. In contrast, a later study by~\citet{Rice2022} based on parametric simulations of the loss of equilibrium of a translationally-invariant flux rope suggested that the helicity ratio was negatively correlated with eruptions. 
Therefore, whether the helicity ratio can be applied to general cases remains controversial. Indeed, all these aforementioned studies are restricted to a particular pre-eruption configuration, i.e., magnetic flux rope that is formed well before the eruption onset (and the eruption is likely initiated by flux-rope instabilities). It remains to see whether similar magnetic parameters can characterize eruption onset caused primarily by magnetic reconnection rather than flux-rope instability, in particular, in the cases with no flux rope existing before eruption~\citep[e.g.,][]{Jiang2021b, Bian2023a}, which are as common as the pre-existing flux rope cases.

In this study, we analyze a set of high-accuracy 3D MHD simulations in which reconnection plays the key role in triggering and driving the eruptions~\citep{Jiang2021, Bian2022a, Bian2022, Bian2023a, Bian2025, Liu2024d, Liu2024c}. 
The simulation set, containing twelve cases in total, is comprehensive in that it includes both bipolar and quadrupolar magnetic configurations, pre-eruption magnetic structures in the form of either sheared arcades or flux rope, and various types of photospheric driving motions, such as shearing, converging, and diffusion. We have previously demonstrated that the initiation of eruption in all cases is governed by a common fundamental mechanism~\citep{Jiang2021b, Jiang2024}. 
Specifically, the slow photospheric motions drive the gradual formation of a core current sheet above the polarity inversion line (PIL), and once magnetic reconnection occurs within this current sheet, the eruption is immediately triggered and is then primarily driven by the ongoing reconnection.
Based on this framework, we systematically investigate the evolution of intensive parameters associated with magnetic energy and helicity in these simulations.


This paper is organized as follows. Section~\ref{sec:data} describes the setup of all the simulations. Section~\ref{sec:methods} introduces the definition and calculation methods of magnetic parameters. Section~\ref{sec:res} presents the evolution of magnetic parameters, including the results for bipolar and quadrupolar configuration, followed by a comparative analysis across all simulations. Finally, summary and discussion are given in Section~\ref{sec:discussion}.

\begin{table*}[!ht]
	\centering
	\caption{Parameters of the Twelve Simulation Cases}
	\begin{tabular}{cccclccccccc}
		\hline
		\multirow{2}{*}{Simulation} & \multicolumn{9}{c}{Parameters}&\multirow{2}{*}{Driving\tablenotemark{\scriptsize 1}} & Eruption \\ 	
		\cline{2-10}
		&$B_c$ &$\sigma_{xc}$&$\sigma_{yc+}$&$\sigma_{yc-}$&$y_c$ &$B_b$ &$\sigma_{xb}$ &$\sigma_{yb}$ &$y_b$ &  &  Onset \\ 
		\hline
		B1 & $32.5$&$2.0$&\multicolumn{2}{c}{$0.5$}&$0.8$&--&--&--&--&R&$78$ \\
		B2 & $20$  &$2.0$&\multicolumn{2}{c}{$1.0$}&$0.8$&--&--&--&--&R&$123$ \\
		B3 & $11.5$&$2.0$&\multicolumn{2}{c}{$2.0$}&$0.8$&--&--&--&--&R&$165$ \\
		B4 & $11.5$&$2.0$&\multicolumn{2}{c}{$2.0$}&$0.8$&--&--&--&--&R+C&$157$ \\
		B5 & $20$  &$2.0$&\multicolumn{2}{c}{$1.0$}&$0.8$&--&--&--&--&R+D&$188.5$ \\
		B6 & $20$  &$2.0$&$0.6$&$0.5$&$0.8$&--&--&--&--&R&$84$ \\
		B7 & $20$  &$2.0$&$0.7$&$0.5$&$0.8$&--&--&--&--&R&$95$ \\
		B8 & $20$  &$2.0$&$0.8$&$0.5$&$0.8$&--&--&--&--&R&$102$ \\
		Q1 & $20$  &$2.0$&\multicolumn{2}{c}{$1.0$}&$0.8$& $10$&$2.0$&$1.0$&$4.0$&R&$93$ \\
		Q2 & $20$  &$2.0$&\multicolumn{2}{c}{$1.0$}&$0.8$& $10$&$2.0$&$1.0$&$6.0$&R&$107$ \\
		Q3 & $20$  &$2.0$&\multicolumn{2}{c}{$1.0$}&$0.8$& $10$&$2.0$&$1.0$&$8.0$&R&$115$\\
		\hline
		   & $R$   &$a$  &\multicolumn{2}{c}{$B_q$}&$L$  &$d$  & & & &&  \\
		TD &$1.8$  &$0.67$&\multicolumn{2}{c}{$63.8$}&$0.35$&$0.65$&--&--&--&C&$28.8$\\
		\hline
	\end{tabular}
	\tablenotetext{1}{R, C, and D represent the rotate, convergence and flux diffusion, respectively.}
	
	\label{tab:define}
\end{table*}

\begin{figure}[!ht]
	\centering
	\includegraphics[width=1.0\textwidth]{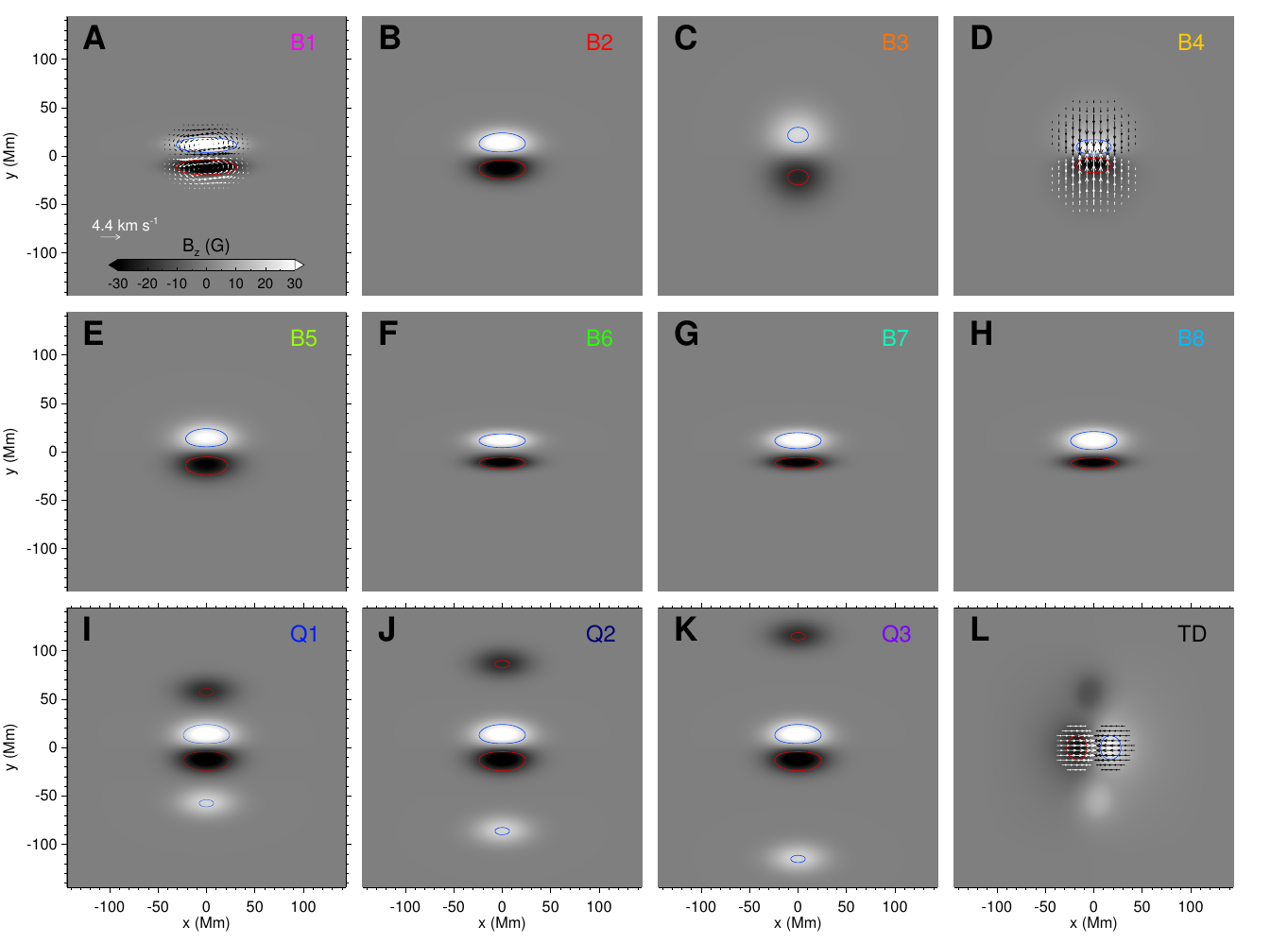}
	\caption{Magnetic flux distributions and photospheric driving flows for all twelve simulation cases. The background shows the 
	vertical magnetic component $B_z$, and the vectors indicate the 
	horizontal surface flows. Blue and red contours mark half of the maximum and minimum $B_z$ values, respectively. Colors correspond to different simulation cases, magenta, red, orange, yellow, light green, green, cyan, light blue, blue, navy blue, purple, and black denote B1, B2, B3, B4, B5, B6, B7, B8, Q1, Q2, Q3, and TD, respectively.} 
	\label{fig:all_map}
\end{figure}

\section{Numerical Model and Simulation Setup}
\label{sec:data}

All simulation data analyzed in this study are taken from a series of previous studies \citep{Bian2022a, Bian2022, Bian2023a, Bian2025, Liu2024d, Liu2024c} that are centered around the reconnection-based CME initiation mechanism \citep{Jiang2021b}. The dataset comprises twelve 3D MHD simulations, the setup parameters of which are given in Table~\ref{tab:define}. Simulations B1 to B5 have a simple sheared arcade in a global bipolar configuration before eruption onset. The shapes of the magnetic polarities differ between B1 and B3, ranging from relatively flat to nearly circular, while B4 and B5 include additional bottom boundary driving.
Simulations B6 to B8 also adopt bipolar configurations, but with varying degrees of asymmetry in positive and negative magnetic flux distributions. 
Simulations Q1 to Q3 have quadrupolar configurations and the core field is a sheared arcade before eruption onset. Variations in the background magnetic field strength lead to different initial heights of the magnetic null point.
The last simulation, TD, contains a flux rope prior to eruption onset.
In this case, the eruption is also triggered and primarily driven by magnetic reconnection. However, unlike the other cases that adopt sheared arcade configurations, the TD case provides a useful control for examining the similarities and differences in the magnetic parameters between systems with and without a pre-existing flux rope.

For the first eleven simulation cases, the initial coronal magnetic field is a potential field constructed from the magnetic flux distribution at the bottom boundary using the Green’s function method. The flux distribution is given by 
\begin{equation}\label{eq:magnetogram}
	\begin{split}
		B_{z}(x,y,0) = B_c
		e^{-x^2/\sigma_{xc}^2}(e^{-(y-y_c)^2/\sigma_{yc+}^2}-e^{-(y+y_c)^2/\sigma_{yc-}^2}),\\
		+ B_b
		e^{-x^2/\sigma_{xb}^2}(e^{-(y-y_b)^2/\sigma_{yb}^2}-e^{-(y+y_b)^2/\sigma_{yb}^2}).
	\end{split}
\end{equation}
where $B_c$ and $B_b$ represent the characteristic strengths of the core and background fields, respectively. Parameters $\sigma_x$ and $\sigma_y$ control the polarity extent in $x$ and $y$ directions, while $y_c$ and $y_b$ specify the separations between the opposite magnetic polarities of the core and background fields. 
In simulations B6 to B8, the two magnetic polarities are assigned different values of $\sigma_{yc}$, with the negative polarity fixed and the positive polarity larger. This leads to a greater extension of the positive polarity  in the $y$ direction and producing an asymmetric magnetic flux distribution.

The initial condition of the last simulation case is constructed using the regularized Biot-Savart law (RBSL) method \citep{Titov2018} combined with an MHD relaxation procedure \citep{Jiang2021b}, yielding an MHD equilibrium state corresponding to Titov-D{\'e}moulin-modified (TDm) model \citep{Guo2019a}. The TDm model is characterized by five independent parameters that specify the geometry and strength of the magnetic flux rope and the background field, the major radius of the flux rope axis ring $R$, the minor radius $a$ of the torus, the strength of the background magnetic charges $B_q$, their sub-photospheric depth $d$, and the half separation between magnetic charges $L$. Table~\ref{tab:define} summarizes the values of parameters used in all simulations.

Fig.~\ref{fig:all_map} displays the photospheric magnetograms and driving flows for all the simulations. In the first eleven simulations, free magnetic energy is injected into the system through an imposed photospheric rotational flow in the core polarities at the bottom boundary, as shown in Fig.~\ref{fig:all_map}A. 
The rotational flow is defined as
\begin{equation}\label{eq:dirven_speed}
	\begin{split}
		v_{x}=\dfrac{\partial \psi(B_{z})}{\partial y}; v_{y}=-\dfrac{\partial \psi(B_{z})}{\partial x},
	\end{split}
\end{equation}
with $ \psi $ given by
\begin{equation}\label{eq:dirven_speed_psi}
	\begin{split}
		\psi = v_{0}B_{z}^{2}e^{-(B_{z}^{2}-B_{z, {\rm max}}^{2})/B_{z, {\rm max}}^{2}},
	\end{split}
\end{equation}
where $B_{z,{\rm max}}$ is the maximum value of the photospheric $B_{z}$, and $v_{0}$ is a constant for scaling such that the maximum of the surface velocity is $4.4$~km~s$^{-1}$.
This incompressible, steady anti-clockwise rotational flow does not modify the flux distribution at the bottom boundary. In cases B1-B3, B6-B8 and Q1-Q3, the rotational driving is maintained throughout the entire run. 
For cases B4 and B5, the rotational flow is terminated at a specific time ($t=140$ for B4 and $t=100$ for B5), well before the eruption onset, and is followed by either a converging flow or a surface magnetic flux diffusion. 
The converging flow drives the opposite polarities toward each other, its velocity profile is defined as follows 
\begin{equation}\label{eq:convergenve}
	\begin{split}
		v_x = 0; v_y = -v_{1}B_z(t=0).
	\end{split}
\end{equation}
Where $v_{1}$ is a constant for scaling such that the largest of the velocity is $4.4$ km s$^{-1}$.
Magnetic flux diffusion is implemented by introducing an artificial resistivity term $\eta_{\text{pho}}$ into the induction equation to mimic photospheric flux diffusion, as given by the following
\begin{equation}\label{eq:mag_induc}
	\dfrac{\partial \vec{B}}{\partial t}= \nabla\times(\vec{v} \times \vec{B}) + \eta_{\text {pho}} \nabla_{\perp}^2 \vec{B}.
\end{equation}
Here, the diffusion coefficient $\eta_{\text{pho}}$ is set to $2 \times 10^{-3}$, corresponding to a characteristic timescale of $500$. In simulation B5, the relaxation phase is imposed from $t=100$ to $150$, followed by the application of surface magnetic flux diffusion.

In the TD case, a converging flow is applied to the strapping field of the flux rope to bring the two main polarities closer together,  as shown in Fig.~\ref{fig:all_map}L, is defined as 
\begin{equation}\label{eq:convergenve_td}
	\begin{split}
		v_x = -v_{2}(B_q)_{z=0}; v_y = 0.
	\end{split}
\end{equation}
Where $v_{2}$ is a constant for scaling such that the largest velocity is $7.4$ km s$^{-1}$. 
It should be noted that all imposed photospheric motions (and the effective flux diffusion) have speed smaller than the Alfv$\acute{\text{e}}$n speed by at least two orders of magnitude, ensuring that the system evolves quasi-statically in response to the surface driving motions.

The MHD equations are solved in a Cartesian coordinate using the conversation-element and solution-element (CESE) scheme~\citep{Feng2010, Jiang2010, Jiang2016a, Jiang2021b}. 
All simulations are performed on adaptive mesh refinement (AMR) grids, with the finest grid resolution ranging from $0.5$ arcsec to $0.125$ arcsec and the coarsest grid of $4$ arcsec.
In all simulation cases, the computational domain is sufficiently large ($768^3$~arcsec$^3$, corresponding to $504^3$~Mm$^3$) to ensure that the simulations are  terminated before any disturbances from the eruption reaches the lateral or upper numerical boundaries.
Hence, the loss of magnetic energy and helicity through the numerical boundaries can be safely neglected. 
In this paper, time is expressed in normalized units with $\tau=105$~s. Further details of the simulation setup can be found in our previous studies~\citep{Jiang2021b, Bian2023a}.
The eruption onset is defined as the time when magnetic (kinetic) energy begins to decrease (increase) rapidly, coinciding with the onset of magnetic reconnection within the core current sheet.

\section{Calculation Method of Magnetic Quantities}
\label{sec:methods}

We investigate evolution of magnetic quantities including the total magnetic energy ($E_{m}$), potential energy ($E_{p}$), and free magnetic energy ($E_{f}$), open field energy ($E_{o}$), relative magnetic helicity ($H_{r}$), and its decomposition into current-carrying ($H_{j}$) and mutual ($H_{pj}$) components. All quantities are evaluated within the entire computational volume. Below we describe the definition of these quantities and the computational method.

\subsection{Magnetic Energies}
\label{sec:em_c}
The total magnetic energy is defined as,
\begin{equation}\label{eq:em}
	E_{m} = \frac{1}{8 \pi}\int_V B^2 \text{d}V,
\end{equation}
where $B$ denotes the magnetic field strength.
The potential magnetic field $\vec{B}_p$, sharing the same normal component as the simulated field on all six boundaries, satisfies
\begin{equation}\label{eq:Bp}
	\nabla \times \vec{B}_{p} = 0, \quad \nabla \cdot \vec{B}_{p} = 0.
\end{equation}
Under these conditions, $\vec{B}_p$ can be derived from a scalar potential that satisfies the Poisson equation.
The corresponding potential field energy is
\begin{equation}
	E_{p} = \frac{1}{8 \pi}\int_V B_p^2 \text{d}V.
\end{equation}
The free magnetic energy represents the deviation of the total magnetic energy from its potential energy, is given by
\begin{equation}
	E_{f} = E_{m} - E_{p}.
\end{equation}

The magnetic energy injected into the coronal volume through photospheric motions can be evaluated using the Poynting flux across the bottom boundary $S$,
\begin{equation}\label{eq:pyt}
	P = \frac{1}{4\pi}\int_S  B_{t}^{2} v_{n}  \text{d}S - \frac{1}{4\pi}\int_S  (\vec B_{t} \cdot \vec v_{t}) B_n
	         \text{d}S,
\end{equation}
where $B_{n}$ and $\vec{B}_{t}$ are the normal and tangential components of the magnetic field, respectively, and $v_{n}$ and $\vec{v}_{t}$ are the corresponding components of the imposed velocity. 
Integrating the Poynting flux in time gives the accumulated energy, denoted as $\Delta E$, 
\begin{equation}\label{eq:ec}
	\Delta E = \int_{0}^{t} P \text{d}t.
\end{equation}

We also computed the open field energy, which represents the upper limit among all simply-connected, force-free fields sharing the same photospheric flux distribution~\citep{Aly1991, Sturrock1991}. The potential and open field energies can be conveniently estimated using the following standard relations,
\begin{equation}\label{eq:energy_pot_open}
	\begin{split}
		E_{p} &= \dfrac{1}{16\pi^{2}} \int_{S\times S'} \dfrac{B_{z}(x,y,0)B_{z}(x',y',0)}{|{\vec{r}}-{\vec{r}}'|}{\rm d}s{\rm d}s', \\
		E_{o} &= \dfrac{1}{16\pi^{2}} \int_{S\times S'} \dfrac{|B_{z}(x,y,0)B_{z}(x',y',0)|}{|{\vec{r}}-{\vec{r}}'|}{\rm d}s{\rm d}s'.
	\end{split}
\end{equation}
This formulation provides a computationally efficient way to estimate $E_p$ and $E_{o}$ directly from the photospheric magnetogram, without solving the 3D Eq.~(\ref{eq:Bp}), with numerical errors below $5\%$~\citep{Bian2022a}. The total magnetic energy is calculated using Eq.~(\ref{eq:em}) with the original AMR grid, the Poynting flux using Eq.~(\ref{eq:pyt}), and the potential and open field energies using Eq.~(\ref{eq:energy_pot_open}).

Furthermore, we use Eq.~(44) of \citet{Valori2016} to evaluate the normalized magnetic energy $E_{\text {div}}/E_{m}$ associated with non-solenoidal component of the magnetic field, which quantifies the effect of a finite numerical divergence of the magnetic field from an energy perspective.
This metric provides a measure of the solenoidal quality and numerical accuracy of the magnetic field in the simulations.

\subsection{Magnetic Helicity}
\label{sec:h_c}
The magnetic helicity within a 3D volume is defined as,
\begin{equation}\label{eq:helicity}
	H = \int_V \vec{A} \cdot \vec{B} \text{d}V.
\end{equation}
where $\vec{A}$ is the vector potential of the magnetic field $\vec{B}$, satisfying $\vec{B}=\nabla\times\vec{A}$.  
The helicity is gauge invariant only when the magnetic field is fully enclosed within the volume $V$. 
Since coronal magnetic fields are generally connected with flux below the photosphere, the relative magnetic helicity~\citep{Finn1985} is used instead,
\begin{equation}\label{eq:rela_helicity}
	H_r = \int_V  (\vec{A}+\vec{A}_p) \cdot (\vec{B}-\vec{B}_p) \text{d}V,
\end{equation}
where $\vec{B}_p$ is the potential field, and $\vec{A}_p$ is its corresponding vector potential. 
This definition of relative helicity is also gauge invariant, and for convenience of computing the 3D vector potential, we follow the method introduced by \citet{Valori2012} using DeVore gauge.

Following \citet{Berger2003}, $H_{r}$ can be decomposed into two components, $H_r=H_{j} + H_{pj}$, in which, 
\begin{equation}\label{eq:H_2}
		\begin{split}
				H_j = \int_V  (\vec{A}-\vec{A}_p) \cdot
				      (\vec{B}-\vec{B}_p) \text{d}V,\\
				H_{pj} = \int_V  2\vec{A}_p \cdot 
				      (\vec{B}-\vec{B}_p) \text{d}V.	
		\end{split}
\end{equation}
Here, $H_{j}$ denotes the helicity of the current-carrying (non-potential) component, and $H_{pj}$ represents the mutual term measuring the coupling between the potential and current-carrying fields.
Since $\vec{B}$ and $\vec{B}_p$ share the same normal component on all six boundaries, all three quantities $H_{r}$, $H_{j}$ and $H_{pj}$ are independently gauge invariant. 
In the following text, $H_{r}$ and $H_{j}$ are referred to as the magnetic helicity and the current-carrying helicity, respectively.
To enable comparisons between different simulations, the square of the photospheric half unsigned magnetic flux $\phi^2$ is used to normalize the helicity with $\phi$ defined as,
\begin{equation}\label{eq:phi}
	\begin{split}
		\phi = \dfrac{1}{2}\int_S |B_z| \text{d}S.	
	\end{split}
\end{equation}

Since magnetic helicity is almost conserved even in the presence of reconnection within the volume, its temporal variation can be estimated from the helicity flux across the bottom boundary~\citep{Berger1984, Berger1999},
\begin{equation}\label{eq:helicity_flux}
	\begin{split}
		\dfrac{\text{d} H_{r}}{\text{d}t} =  2 \int_S (\hat{\vec{A}}_{p} \cdot \vec{B}_{t}) v_{n} \text{d} S - 2 \int_S (\hat{\vec{A}}_{p} \cdot \vec{v}_{t}) B_{n} \text{d} S.
	\end{split}
\end{equation}
Note that this formula of helicity flux requires a particular gauge for the vector potential $\hat{\vec{A}}_{p}$, which is $\nabla \cdot \hat{\vec{A}}_{p} = 0$, and $\hat{A}_{pz}=0$~\citep{Berger1999, Schuck2008, Pariat2015}. Therefore $\hat{\vec{A}}_{p}$ is different from the vector potential $\vec A_p$ in Eq.~(\ref{eq:rela_helicity}). 
If there is no loss of helicity from the side and top boundaries and if no dissipation in the volume, the accumulated helicity obtained from this flux integration should match the volume helicity computed using Eq.~(\ref{eq:rela_helicity}). This consistency provides a useful diagnostic for assessing the level of helicity dissipation in the simulations~\citep{Yang2013b, Pariat2015}.


\subsection{Influence of Numerical Resolution on Magnetic Helicity Calculation}
\label{sec:h_res}

Since the original simulations were performed using AMR, the magnetic field data need to be interpolated onto uniform grids to facilitate the calculation of magnetic helicity based on the volume integration (Eqs.~\ref{eq:rela_helicity} and \ref{eq:H_2}). However, it is prohibitive to perform the interpolation and integration using the original highest resolution in each simulation. For example, in simulation case B2, the entire computational volume is $700^3$~arcsec$^3$ and if we consider to use a resolution of $0.5$~arcsec (which is the original highest resolution in the AMR grid used in this case), the total number of grid points would be $1400^3$ ($\sim 2.7$ billion), requiring about $192$~GB memory, which would be prohibitively expensive for long-term, time-dependent MHD simulation data. Thus we choose an unified, relatively low resolution that we can afford for computations of all the different simulation cases. In all simulations presented in this study, a uniform grid resolution of $2$ arcsec was adopted as a compromise between computational efficiency and numerical resolution. It is essential to evaluate how the degradation of grid resolution affects the numerical accuracy of magnetic helicity. Here we used the simulation case B2 as an example.

\begin{figure}[!htb]
	\centering
	\includegraphics[width=0.8\textwidth]{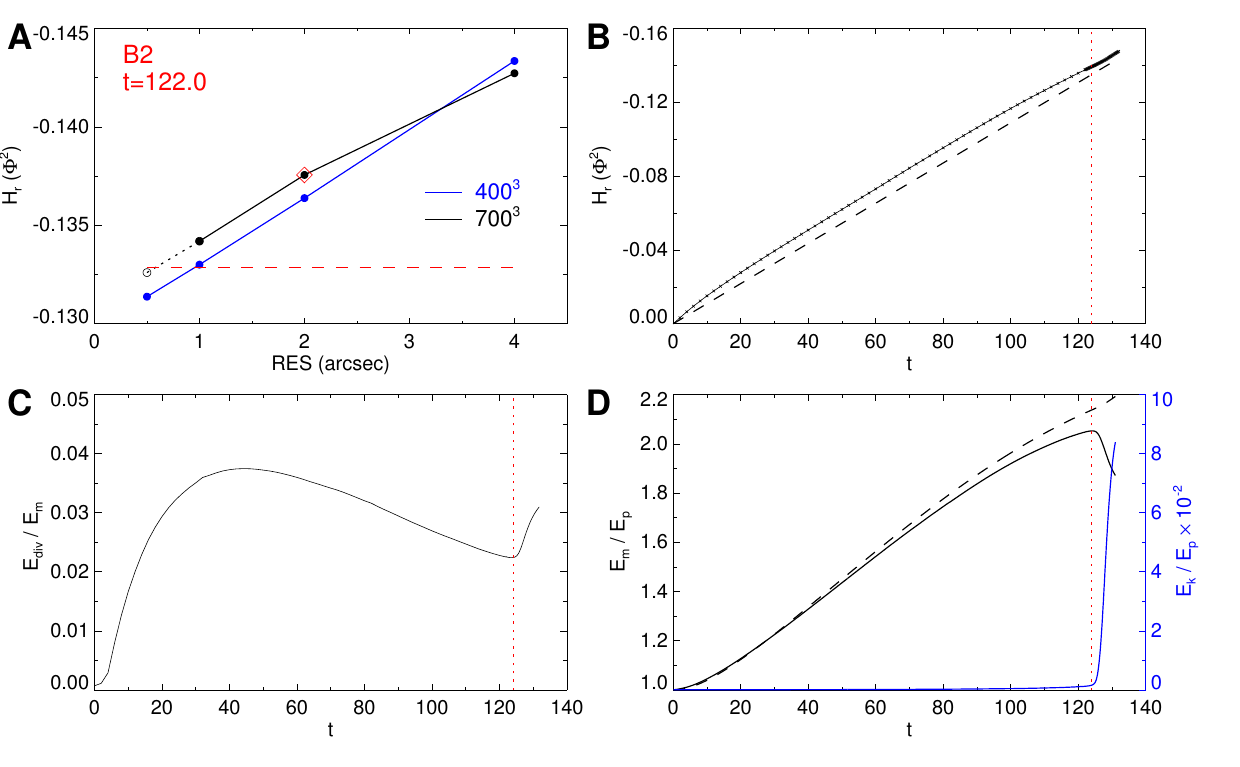}
	\caption{Magnetic helicity calculation and verification of the magnetic field accuracy in simulation case B2.
		(A)	Dependence of magnetic helicity on spatial resolution and computational domain in simulation case B2 at $t=122$.
		Each dot represents a calculation performed at uniform grid resolution of $0.5$, $1$, $2$, and $4$ arcsec. At the highest resolution of $0.5$ arcsec, only the result within the central region of $400^3$ arcsec$^3$ is available due to the limitation of computational resources. The hollow point denotes value obtained by linear extrapolation from the low resolution data. The red diamond denotes the values adopted in helicity calculation. The red horizontal dashed line represents the accumulated helicity injected from the bottom boundary at a grid resolution of $0.5$ arcsec. 
		(B) temporal evolution of magnetic helicity at a computational domain of $700^3$ arcsec$^3$. The solid cross line denotes the total magnetic helicity at a grid resolution of $2$ arcsec, with each cross denoting an output snapshot. The dashed line indicates the helicity flux injected through the bottom boundary at a grid resolution of $0.5$ arcsec. The vertical red dotted line marks the eruption onset.	
		(C) temporal evolution of the normalized magnetic energy associated with non-solenoidal field $E_{div}/E_{m}$.
		(D) temporal evolution of magnetic energy (black) and kinetic energy (blue) in the simulation, with the black dashed line showing the Poynting flux injected from the bottom boundary.} 
	\label{fig:h_res}
\end{figure}

Fig.~\ref{fig:h_res}A presents the dependence of magnetic helicity on the spatial resolution (from a lowest resolution of $4$~arcsec to a highest one of $0.5$~arcsec). The helicity is computed for time of $t=122$, which is slightly before the eruption onset. As a reference value~(indicated by the red dashed line), the time integrated helicity injection from the bottom boundary is computed with Eq.~(\ref{eq:helicity_flux}), for which we can use the original highest resolution in the simulation. The black dots represent results obtained for the full simulation box ($700^3$~arcsec$^3$), for which we can only compute with resolutions up to $1$~arcsec. For comparison, we also attempted to compute with resolutions up to $0.5$~arcsec, but for a central sub-volume ($400^3$~arcsec$^3$), which is shown by the blue dots. 
For a given resolution, the full volume value ought to be slightly higher than the sub-volume one, unless the numerical error becomes evident. As can be seen, this is indeed fulfilled for the resolutions of $1$ and $2$ arcsec, but not for the resolution of $4$ arcsec, suggesting that the $4$ arcsec result has somewhat larger errors. As the grid resolution becomes lower, the helicity increase systematically, showing a nearly linear relationship between the computed values and the grid resolution. This allows us to estimate the full-volume helicity at the highest spatial resolution of $0.5$ arcsec through linear extrapolation, which is shown by the hollow point in Fig.~\ref{fig:h_res}A. It can be seen that the volume integration result at the highest resolution match pretty well with the accumulated helicity obtained from the boundary injection, which demonstrates the high fidelity of the simulation.

Fig.~\ref{fig:h_res}B compares the temporal evolution of the magnetic helicity computed by volume integration with resolution of $2$ arcsec (the solid line) and that by time integration of helicity injected from the photospheric boundary using the original highest resolution (the dashed line). Note that the helicity injection rate is a constant according to Eq.~(\ref{eq:helicity_flux}), since for this simulation case, $v_n=0$, therefore the first integration term is zero, and all the variables (i.e., $\vec{v}_{t}$, $B_{n}$ and $\vec{A}_{p}$) in the second integration term do not change with time. Due to the resolution degradation effect, the volume integrated helicity is always slightly larger than the value derived from the surface injection, but the their changing rates are consistent with each other. 
In Fig.~\ref{fig:h_res}C, we show the temporal evolution of the normalized magnetic energy $E_{\text{div}}/E_{m}$ associated with the non-solenoidal component of the magnetic field in case B2. This value remains below $4\%$ throughout the simulation, indicating good solenoidal quality of the magnetic field and supporting the reliability of the helicity calculation.
Fig.~\ref{fig:h_res}D shows that the magnetic energy computed from the simulation volume (with on the original AMR grid) agrees well with the accumulated energy injected through the bottom boundary, with only a slight deviation before the eruption, since a small part of the magnetic energy is released for the slow-expansion of the pre-eruptive field. Before the eruption, both magnetic helicity and energy increases continuously, but after eruption they behavior very differently. Notably, the growth of relative magnetic helicity is almost unaffected by the eruption, consistent with the theoretical expectation that helicity is almost conserved (and numerical dissipation is weak), whereas the magnetic energy undergoes a rapid release once the eruption begins, with part of the energy converted to the kinetic energy which therefore shows an impulsive increase.

\section{Results}
\label{sec:res}

In the section, we first examine two typical simulations and then perform a comparative analysis across all twelve simulations.

\begin{figure}
	\centering
	\includegraphics[width=0.8\textwidth]{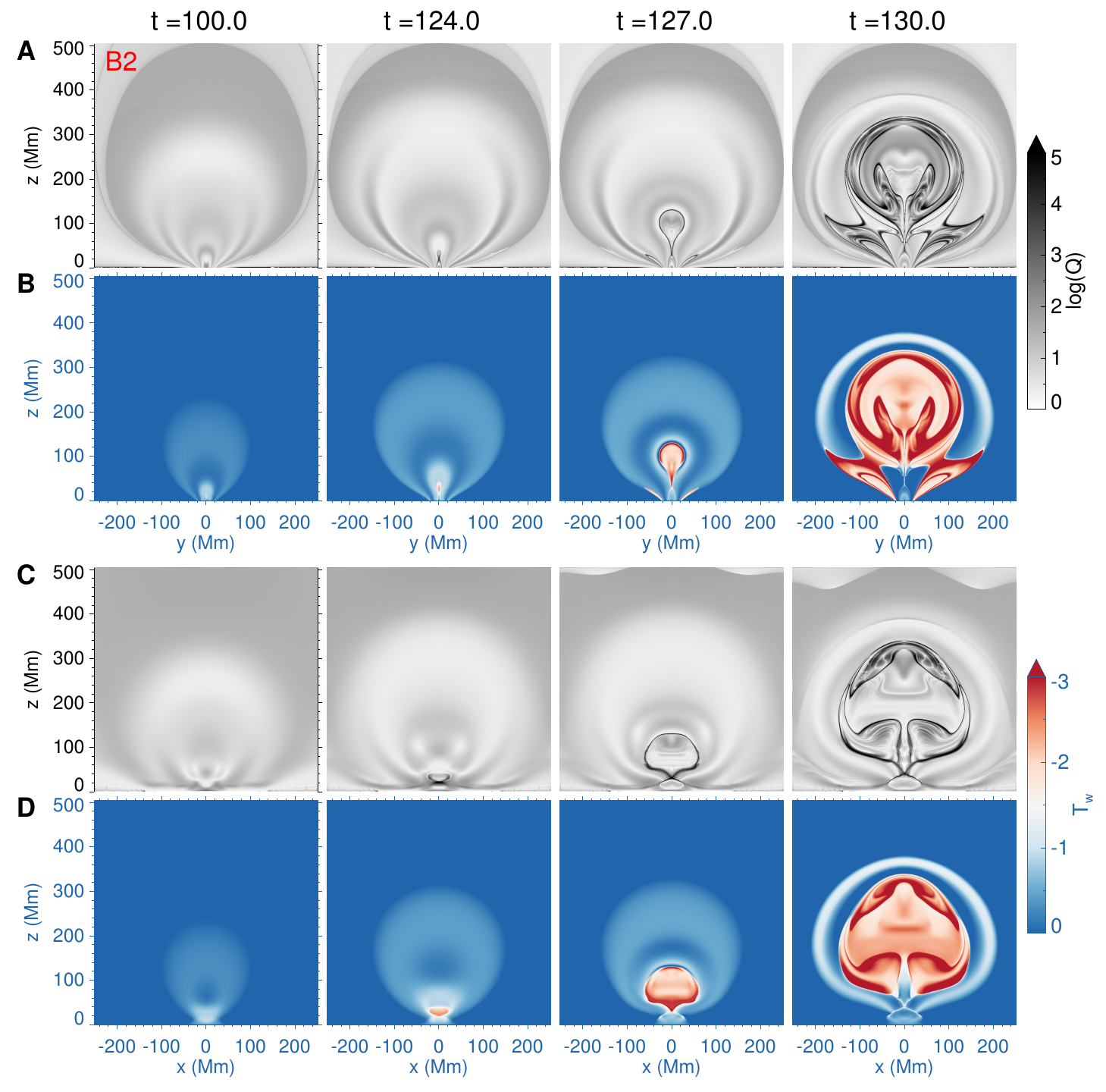}
	\caption{Evolution of the magnetic topology in simulation case B2. Panels (A) and (B) show the squashing factor $Q$ and the  twist number $T_{w}$ on the $x=0$ slice, while (C) and (D) show the same quantities on the $y=0$ slice.} 
	\label{fig:qsl_B2}
\end{figure}
\begin{figure}
	\centering
	\includegraphics[width=0.8\textwidth]{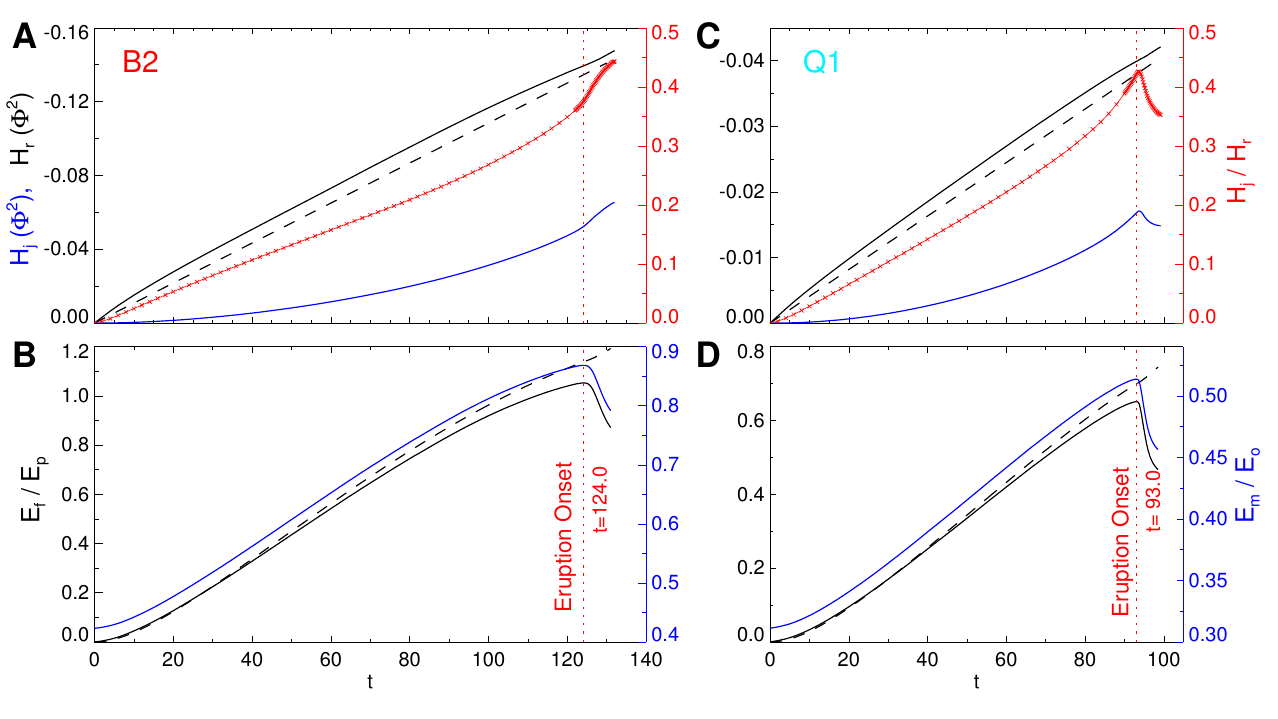}
	\caption{Temporal evolution of the normalized magnetic helicity and energies for simulation cases B2 and Q1. 
	(A) shows the temporal evolution of magnetic helicity at a computational domain of $700^3$ arcsec$^3$. 
	The solid line denotes the total magnetic helicity at a grid resolution of $2$ arcsec, and the dashed line indicates the helicity flux injected through the bottom boundary at a grid resolution of $0.5$ arcsec. The blue line denotes the current -carrying helicity. The red cross line represents the helicity ratio $H_j/H_r$, with each cross denoting an output snapshot. The vertical red dotted line marks the eruption onset.
	(B) shows the normalized free magnetic energy $E_{f}/E_{p}$ (black) and total magnetic energy $E_{m}/E_{o}$ (blue).
	(C) and (D) show the corresponding quantities for simulation case Q1.}
	\label{fig:h_Q1_B2}
\end{figure}

\subsection{Simulation of Bipolar Configuration B2}
\label{sec:b2}

Simulation B2 begins with a single bipolar magnetic field driven by continuous photospheric rotational flow. The coronal magnetic field evolves from an initial potential field into a highly sheared configuration above the PIL, where the current density becomes strongly enhanced. A vertical current sheet forms at $t=124$, while the surrounding field remains nearly current-free. Once the current sheet thins to the grid resolution, magnetic reconnection sets in and triggers the eruption. This is marked by a sharp release of magnetic energy and a sharp rise in kinetic energy. Fig.~\ref{fig:qsl_B2} illustrates the magnetic topology evolution, characterized by the magnetic squashing factor $Q$ and the twist number $T_w$ on the two central vertical slice, i.e., the $x=0$ and $y=0$ planes. The squashing factor $Q$ can be used to quantify the magnetic connectivity. 
The high $Q$ values indicate regions where magnetic field lines diverge significantly, corresponding to sites where 3D magnetic reconnection preferentially occurs. The boundaries of distinct magnetic domains, where $Q=\infty$, are known as separatrices, while quasi-separatrix layers (QSLs) corresponding to regions with $Q\approx2$. The twist number $T_w$ quantifies the number of turns that two infinitesimally close field lines wind around each other~\citep{Berger2006}.

The temporal evolution of helicity and energy is presented in Fig.~\ref{fig:h_Q1_B2}A-B. To account for differences in magnetic flux at the bottom boundary among different simulation cases, the helicity is normalized by the square of half the unsigned magnetic flux at the bottom boundary, while the magnetic energies are normalized by their potential energy and open field energy, respectively. 
Under the constant injection of helicity flux from the bottom boundary, the magnetic helicity $H_r$ (black solid line), computed by volume integration at a resolution of $2$ arcsec, increases linearly. Fig.~\ref{fig:h_Q1_B2}A also shows the time integrated helicity injected through the photospheric boundary using the original highest resolution (dashed line). According to Eq.~(\ref{eq:helicity_flux}), the helicity injection rate remains constant. In this simulation case, $v_n=0$ , therefore the first term vanishes while the second term does not change with time. 
Owing to resolution degradation effect, the volume integrated helicity is systematically slightly larger than the value derived from the surface injection, but their changing rates are consistent with each other.
The current-carrying helicity $H_j$ (blue line) shows a growing increasing rate before the eruption onset, and once the eruption begins, the growth rate is sharply accelerated. This behavior is more clearly reflected in the evolution of the helicity ratio $H_j/H_r$ (red line), which reaches $0.35$ just prior to the eruption and then rises more rapidly afterward.
This rapid acceleration is attributed to the magnetic reconnection within the current sheet during the eruption, which transforms the strongly sheared arcades into a twisted flux rope (see Fig.~7A in \citet{Bian2022a}), thereby converting part of the potential field component into the current-carrying component. Fig.~\ref{fig:h_Q1_B2}B shows the temporal evolution of the two sets of normalized energies. Before the eruption, both the total magnetic energy $E_{m}/E_{o}$ and free energy $E_{f}/E_{p}$ increase continuously and is followed by a rapid decrease at the eruption onset. At the onset time, $E_{f}/E_{p}$ reaches a maximum value of approximately $1.0$, indicating that the free magnetic energy becomes comparable to the potential field energy; the total magnetic energy corresponds to about $0.87$ of the open field energy, suggesting that the system is close to the open field state. 

Before the eruption, both magnetic helicity and energy increases continuously, but after eruption they behavior very differently. Notably, the growth of relative magnetic helicity is almost unaffected by the eruption, consistent with the theoretical expectation that helicity is almost conserved, whereas the magnetic energy undergoes a rapid release once the eruption begins, with part of the energy converted to the kinetic energy which therefore shows an impulsive increase.

\begin{figure}
	\centering
	\includegraphics[width=0.8\textwidth]{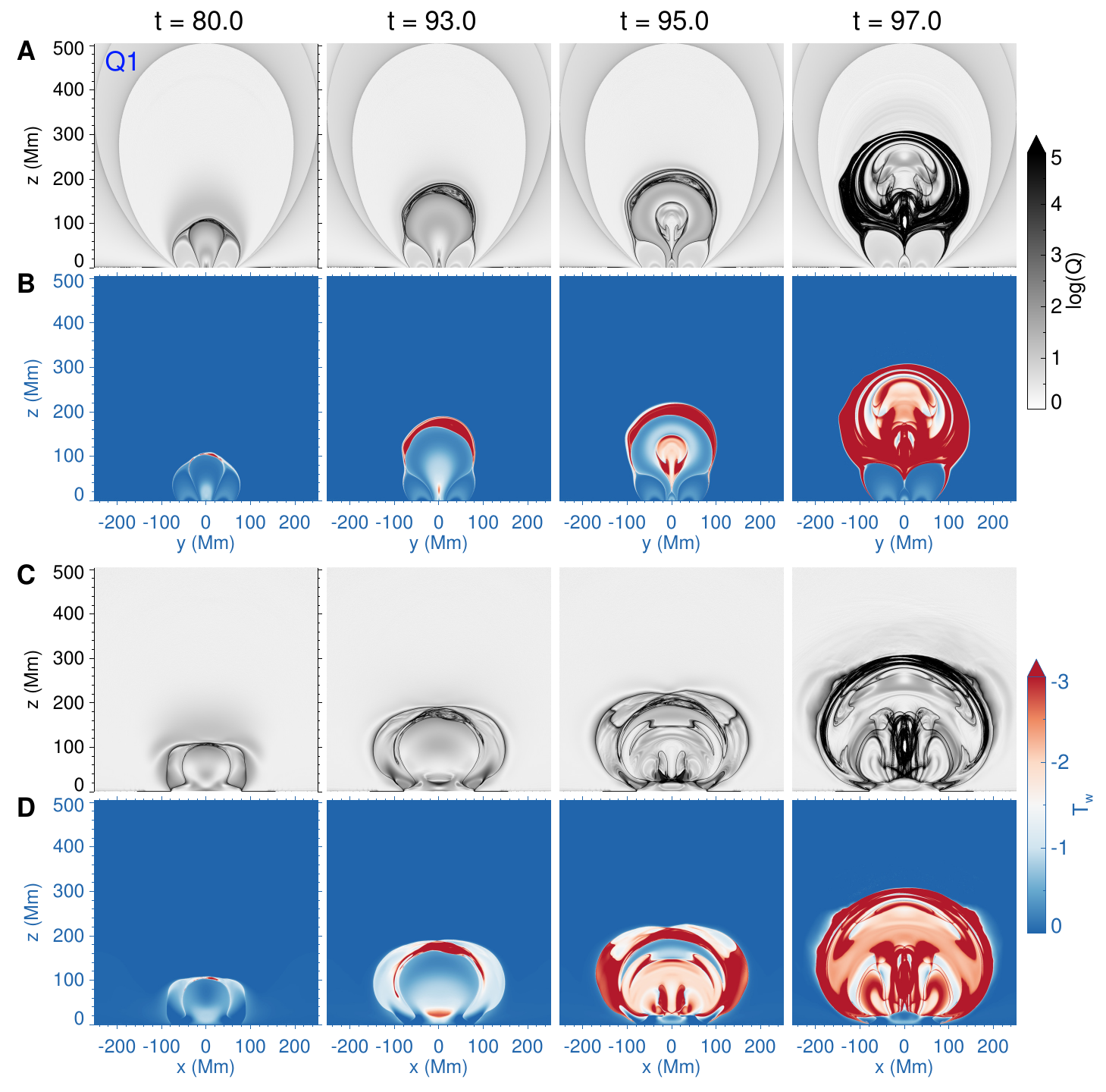}
	\caption{Evolution of the magnetic topology in simulation case Q1. Panels (A) and (B) show the squashing factor $Q$ and the magnetic twist number $T_{w}$ on the $x=0$ slice, while (C) and (D) show the same quantities on the $y=0$ slice.} 
	\label{fig:qsl_Q1}
\end{figure}

\subsection{Simulation of Quadrupolar Configuration Q1}
\label{sec:q1}

Simulation Q1 has a quadrupolar magnetic configuration, of which the initial magnetic field contains a null point above the central bipolar field (Fig.~\ref{fig:qsl_Q1}). Under photospheric rotational driving, the core field expands, compressing the null point and forming a breakout current sheet. At $t=80$, reconnection has already taken place within the breakout current sheet, whereas the underlying core field remains in quasi-static expansion. During this phase, magnetic energy increase steadily, while kinetic energy remains negligible. With continued driving, the breakout current sheet elongates further and the current distribution in the core field becomes progressively compressed into a vertical current sheet, forming around $t=93$. Once the core current sheet thins to the grid resolution, reconnection kicks in there, resulting in a sharp release of magnetic energy and a rapid rise in kinetic energy, marking the eruption onset.

The temporal evolution of helicity and energy is presented in Fig.~\ref{fig:h_Q1_B2}C-D. Under the driving of photospheric rotational flows, the total magnetic helicity $H_r$ (black line) exhibits a continuous linear increase, consistent with the helicity flux injected through the bottom boundary. Similar to the bipolar configuration B1, before the eruption onset, the current-carrying helicity $H_j$ increase with growing rate, but after the eruption onset, its behavior is totally different; it first rises slightly (only within a very short period) and then drops rapidly. This behavior is also mirrored in the helicity ratio $H_j/H_r$, which reaches its maximum ($\sim 0.42$) shortly after eruption onset and then decreases sharply. This decline is caused by the breakout reconnection between the erupting flux rope and the external field, which transforms part of the current-carrying field lines of the rope into simpler magnetic arcades on both sides (Fig.~\ref{fig:qsl_Q1}C-D). Fig.~\ref{fig:h_Q1_B2}D shows the two sets of normalized magnetic energies. They exhibits continuously increase before the eruption onset and a sharp decline afterwards. At the eruption onset, the free magnetic energy $E_f/E_p$ reaches about $0.6$, and the total magnetic energy $E_m/E_{o}$ is roughly $0.5$, implying that the entire system is still far away from the open field state when the eruption begins.

\begin{figure}
	\centering
	\includegraphics[width=1.0\textwidth]{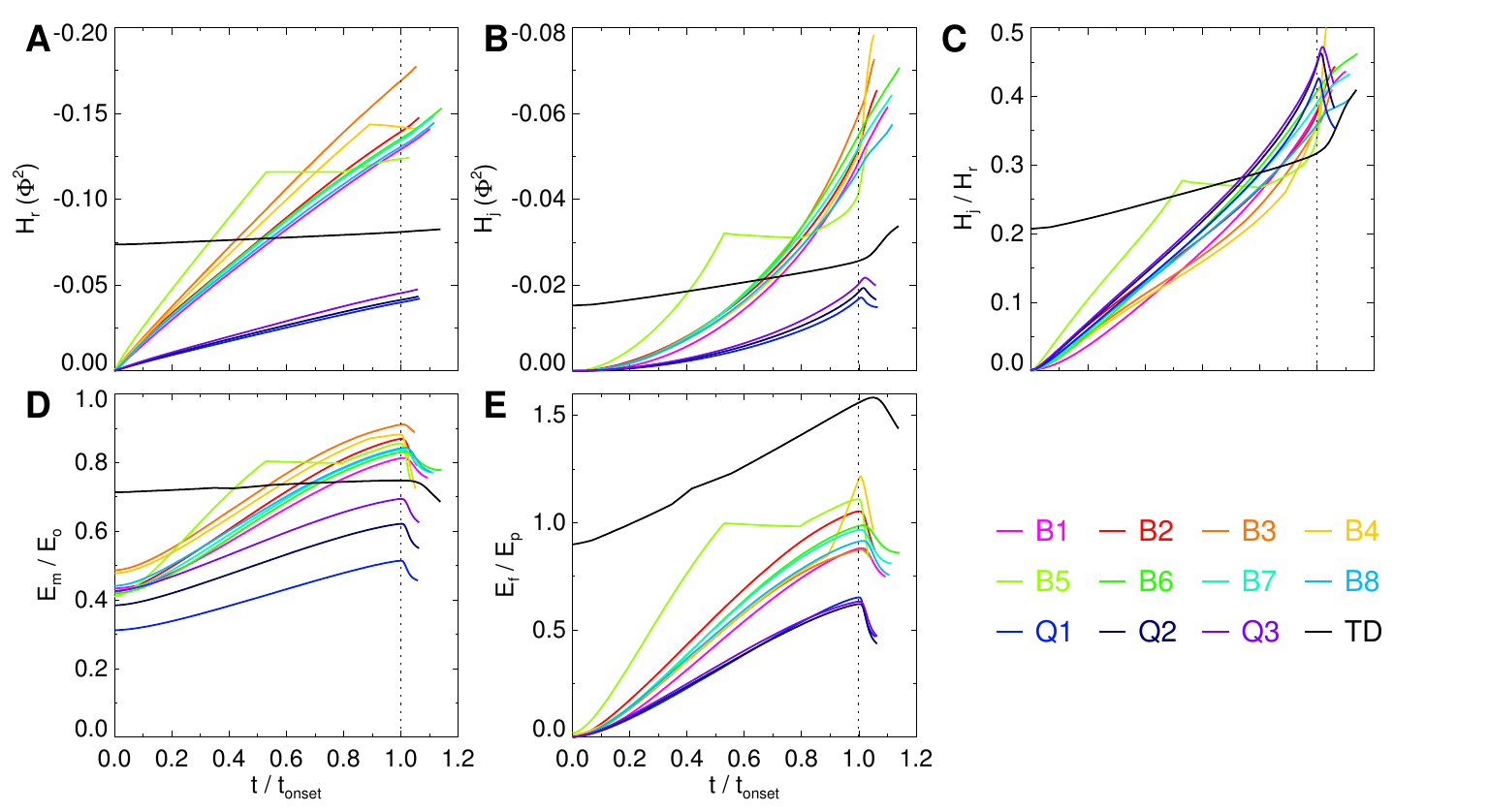}
	\caption{Temporal evolution of the normalized physical quantities for all twelve simulations. Panels (A)–(E) show the normalized magnetic helicity $H_r/\Phi^2$, normalized current-carrying helicity $H_j/\Phi^2$, helicity ratio $H_j/H_r$, normalized total magnetic energy $E_m/E_o$, and normalized free energy $E_f/E_p$, respectively. 
	The time axis is normalized by the eruption onset time of each simulation case, such that $t=1$ corresponds to the eruption onset (vertical dashed line).} 
	\label{fig:h_all}
\end{figure}

\subsection{Comparison Across Twelve Simulations}
\label{sec:all}

Fig.~\ref{fig:h_all} summarizes the evolution of the energy and helicity quantities for all the twelve simulations. The time axis is normalized by the eruption onset time for each simulation case, providing a unified view of the temporal evolution. Therefore, here $t=1$ denotes the eruption onset in each case.

As shown in Fig.~\ref{fig:h_all}A, for all simulation cases, the onset of eruption (i.e., the start of reconnection) has virtually no impact on the evolution of $H_r$. This is consistent with theory that magnetic helicity remains highly conserved even in the presence of reconnection within the volume, and therefore the eruption does not change the $H_r$ evolution. It also demonstrates that our simulations have very low numerical resistivity. 
$H_r$ starts from zero (corresponding to potential field) for all cases except TD, where the pre-existing twisted flux rope has a nonzero initial relative helicity. Moreover, the quadrupolar cases display smaller normalized helicities compared with bipolar ones, since a substantial fraction of their magnetic flux belongs to the background potential field, which carries no helicity. However, the temporal evolution of the current-carrying helicity $H_j$ differs markedly between the bipolar and quadrupolar configurations, as shown in Fig.~\ref{fig:h_all}B. All bipolar cases, including the TD, exhibit a pronounced acceleration in the helicity growth immediately after the eruption onset. In contrast, all quadrupolar cases show a very short rise followed by a rapid decline after the eruption onset time. Before eruption, the increase of $H_j$ results from the continuous helicity flux injection through the bottom boundary. With start of eruption, the evolution of $H_j$ results from not only the ongoing bottom boundary injection but also magnetic reconnection. The reconnections at different locations have different effects on the evolution of $H_j$; reconnection in core current sheet transforms strongly sheared arcades into twisted flux ropes, which increases $H_j$, but in the quadrupolar case, the external breakout reconnection will peel off part of the flux from the flux rope, which reduces $H_j$. Therefore, before the eruption, the injection rate dominates over current-carrying helicity loss, leading to a net increase. After the eruption onset, both core and breakout reconnection operate simultaneously, as the system expands, breakout reconnection becomes more efficiently at removing current-carrying helicity, resulting in a brief increase followed by a sharp decrease.

The evolution of the helicity ratio $H_j/H_r$, as shown in Fig.~\ref{fig:h_all}C, exhibits a similar trend to the that of $H_j$. In all bipolar and the TD cases, the ratio continues to rise after the eruption onset, while in all quadrupolar cases, it briefly increases and then rapidly drops. This behavior depends on whether the core or breakout reconnection dominates the changing of the magnetic topology of erupting field. Interestingly, all simulations have a helicity ratio $H_j/H_r$ close to $0.4$ at eruption onset, whereas such a consistence is not seen in the evolution of either $H_j$ or $H_r$.

As shown in Fig.~\ref{fig:h_all}D, the normalized magnetic energy increases before the eruption and decreases afterward in all simulations, spanning a range of $0.5$ to $0.9$. In bipolar cases, the normalized magnetic energy exceeds $0.8$ at eruption onset, whereas in quadrupolar cases it remains below $0.7$. 
This indicates that the bipolar cases requires a substantially higher overall degree of non-potentiality to reach an eruptive state. In contrast, the quadrupolar cases exhibit a comparatively lower global non-potentiality, since only the central core region needs to form a current sheet. Similarly, Fig.~\ref{fig:h_all}E shows that the normalized free magnetic energy exhibits the same overall trend, bipolar cases possess stronger non-potentiality ($>0.8$) at eruption onset, while quadrupolar cases remain around $0.6$. The TD case is not directly comparable, as its imposed photospheric driving rapidly decreases the potential energy while keeping the open field energy nearly unchanged, leading to an artificially steep rise in the normalized free magnetic energy prior to the eruption.

\begin{table}[!ht]
	\centering
	\caption{Magnetic Quantities of the Twelve Simulations at Eruption Onset.}
	\begin{tabular}{lccccc}
		\hline
		 & $H_r (\phi^2)$ & $H_j (\phi^2)$ & $H_j/H_r$ & $E_m/E_o$ & $E_f/E_p$ \\ 	 
		\hline
		Mean & $0.1127$ & $0.0409$ & $0.3834$ & $0.7843$ & $0.9530$ \\
		SD\tablenotemark{\scriptsize 1} & $0.0441$ & $0.0150$ & $0.0405$ & $0.1285$ & $0.2659$ \\
		COV\tablenotemark{\scriptsize 2}& $39.15\%$ & $36.74\%$ & $10.57\%$ & $16.38\%$ & $27.91\%$ \\
		\hline
	\end{tabular}
	\tablenotetext{1}{Standard deviation.}
	\tablenotetext{2}{Coefficient of variation.}
	\label{tab:para}
\end{table}

We further studied the distribution of various magnetic quantities at the eruption onset across twelve simulations, calculating their mean values, standard deviations, and coefficients of variation, as shown in Table~\ref{tab:para}. It can be seen that the helicity ratio $H_j/H_r$ has the most consistent values of $0.38\pm0.04$ at the eruption onset, with a coefficient of variation (COV) of around $10\%$. This indicates the robustness of the helicity ratio as an indicator of eruption onset, at least in our compared simulations. The $H_r$ and $H_j$ have a wide range with COV of $35\%\sim40\%$. The energy indicators have smaller COV, in particular, the $E_m/E_o$ has values of $0.78\pm0.13$ with COV of around $16\%$.

\section{Conclusion and Discussions}
\label{sec:discussion}

In this study, based on a range of numerical MHD models of solar eruption initiation with different coronal magnetic topologies and different driving motions at the bottom surface (photosphere), we have systematically examined the evolution of intensive parameters associated with magnetic energy and helicity, aiming to seek threshold-like parameters with ability to predict the eruption onset in various cases. These numerical models are adopted from our previous studies, and they include both simple bipolar configurations and complex quadrupolar ones, with both sheared arcade and MFR before eruption onset, and with various surface motions such as shearing, rotation, converging, and diffusion motions. Nevertheless, we have previously demonstrated that the initiation of eruption in all the different cases is governed by a fundamental mechanism. That is, the formation of a current sheet above the core PIL plays a crucial role in leading to eruptions, and once magnetic reconnection occurs within this current sheet, the eruption is immediately triggered and is primarily driven by the reconnection.

Before examining the parameters, we have carefully checked the computing accuracy of both magnetic helicity and energy, which shows that the simulations have a sufficiently high fidelity. The intensive parameters being analyzed include the normalized relative helicity $H_r/\phi^2$, the normalized current-carry helicity $H_j/\phi^2$, the ratio of current-carrying helicity to relative helicity $H_j/H_r$, the ratio of total magnetic energy to open-field energy $E_m/E_o$, and the ratio of free magnetic energy to potential field energy $E_f/E_p$. All these parameters measure the degree of non-potentiality of the coronal field but with different aspects. All these parameters show a overall increase before the eruption, owing to the persistent driving motions that inject helicity and free energy into the system. In contrast, the specific values at the eruption onset as well as their evolution after eruption behavior quite differently.

Among all examined parameters, it is found that the helicity ratio $H_j/H_r$, as first proposed by \citet{Pariat2017}, exhibits the most consistent threshold behavior that can more likely indicate the eruption onset in all the simulation cases than other parameters. Specifically, at the eruption onset, $H_j/H_r$ converges to a narrow range of approximately $0.38\pm0.04$ across all simulations (with a COV of around $10\%$), regardless of the different magnetic topologies and bottom driving motions. Strikingly, this finding of $H_j/H_r$ with eruption threshold of around $0.38$, is close to the value of $0.45$ that can discriminate eruptive cases from non-eruptive ones as found by \citet{Pariat2017} in simulations of twisted flux rope emergence into background fields of different strengths and orientations. It is also close to the value $0.3$ as found by \cite{Zuccarello2018} with a different set of simulations in which all eruptions are initiated by torus instability of a pre-existing MFR. To find a theoretical explanation of the threshold-like behavior of $H_j/H_r$, both \citet{Pariat2017} and \cite{Zuccarello2018} suggest that this helicity ratio may be related to torus instability, i.e., a proxy of the instability criterion. Such a framework of interpretation is distinct from the situation in our simulations, in which (except the TD case) there is no MFR but sheared arcade before eruption. Therefore the helicity ratio cannot discriminate the different initiation mechanisms.


Interestingly, the evolution of the helicity ratio after the eruption onset behaviors distinctly between the two types of magnetic configurations, which has not been seen in previous studies. In the bipolar cases (including the TD one), reconnection within the sheared core arcade increases the current-carrying helicity, leading to a monotonic rise of $H_j/H_r$ after the eruption. In contrast, in the quadrupolar cases, the helicity ratio first increases and then rapidly decreases shortly after eruption onset, reflecting the competing effects between the core current sheet reconnection and the breakout reconnection (which decreases $H_j/H_r$).



These results collectively highlight the helicity ratio as a robust diagnostic parameter for assessing the eruptive potential of coronal magnetic fields. At eruption onset, the helicity ratio converges to a relatively narrow critical range (for example, $0.38\pm0.04$ in our studies and $0.3$ as found in~\citet{Zuccarello2018}), largely independent of magnetic topology and eruption mechanism.
This relatively narrow range reflects the particular conditions at eruption onset rather than the subsequent temporal evolution, emphasizes its potential value for physics-based eruption prediction. 
Compared with observational studies, the critical values obtained here are systematically higher than the typical observational results of about $0.1$ reported by \citet{Moraitis2019}, \citet{Gupta2021}, and \citet{Thalmann2025}. This difference is likely related to the idealized nature of the simulations, in which most of the magnetic flux distribution coherently participates in the eruption process, whereas in active regions only part of the magnetic flux system is directly involved and different photospheric motions may inject helicity of opposite signs, thereby reducing the net current-carrying helicity. 
Furthermore, the distinct post-eruption evolution patterns of $H_j/H_r$ provide new insight into the competing roles of different types of magnetic reconnection in shaping magnetic helicity evolution. 
In future work, we plan to extend the parameter space of the simulations to include a wider range of boundary conditions, driving profiles, and flux imbalances, thereby exploring how these factors influence the helicity threshold. Combining the helicity ratio with other topological and current related parameters, such as current helicity and magnetic twist parameters, will help to establish more precise and physically grounded criteria for distinguishing between eruptive and confined flares. Ultimately, identifying universal and topology independent eruptivity thresholds for solar eruptions represents a important step toward development of physics-based prediction models and the improvement of space weather forecasting capabilities.

\begin{acknowledgments}
	This work is jointly supported by National Key R\&D Program of China under No. 2024YFA1612001, National Natural Science Foundation of China (NSFC 42504160, 12573058), Shenzhen Science and Technology Program (grant no. RCJC20210609104422048), Guangdong Basic and Applied Basic Research Foundation (2023B1515040021), and the Specialized Research Fund for State Key Laboratory of Solar Activity and Space Weather. 
\end{acknowledgments}


%
\bibliography{mylib}  
\bibliographystyle{aasjournal}

\end{document}